\documentclass[twocolumn,amsmath,amssymb,aps,pra,10pt,showkeys
,longbibliography
,nofootinbib
]{revtex4-2}
\usepackage{dcolumn}
\usepackage{bm}
\usepackage{mathtools,cases}
\usepackage[unicode]{hyperref}
\usepackage{hyperref}
\usepackage{graphicx}
\usepackage{times}
\usepackage{braket}
\hypersetup{colorlinks=true,linkcolor=blue,citecolor=blue,filecolor=black,urlcolor=blue}

\makeatletter

\begin{document}

\title{The variational theorem for the scattering length in low dimensions and its applications to universal systems}
\author{Alexander Yu.~Cherny}
\email{cherny@theor.jinr.ru}
\affiliation{Bogoliubov Laboratory of Theoretical Physics, Joint Institute for Nuclear
Research, 141980, Dubna, Moscow region, Russia}

\date{\today}

\begin{abstract}
The variational theorem for the scattering length \cite{cherny00} is extended to one and two dimensions. It is shown that the arising singularities can be treated in terms of generalized functions. The variational theorem is applied to a universal many-body system of spinless bosons. The extended Tan adiabatic sweep theorem is obtained for interacting potentials of arbitrary shape with the variation of the one-particle dispersion. The pair distribution function is calculated at short distances by means of the variation of the potential. The suggested scheme is based on simple quantum mechanics; it is physically transparent and free from any divergence.
\end{abstract}

\maketitle

\section{Introduction}
\label{sec:intro}

For cold atoms, interactions between particles are governed by the low-momentum $s$-wave scattering amplitude, which is determined by the scattering length \cite{Leggett01,book:Pethick08,Pitaevskii16:book}. This property is called universality. When the scattering length becomes very large, many-body systems are close to the unitary regime, where the mean-field theoretical description becomes irrelevant and other approaches are needed \cite{Braaten12,Chevy16}. A phenomenological thermodynamic approach is to consider the scattering length, which can be controlled with the Feshbach resonance, as a thermodynamic property. The Tan adiabatic sweep theorem \cite{Tan2008b,Tan2008a} relates the increment of energy to infinitesimal change of the scattering length when it changes slowly.

In the previous publication \cite{cherny21}, we derived the Tan theorem for a system of spinless bosons in three dimension (3D) by means of the variational theorem for the scattering length \cite{cherny00}. In this paper, we generalize the variational theorem to lower dimensions and apply it to the Tan theorem in 1D and 2D. Besides, we calculate the short-range part of the pair distribution function in these dimensions.

The 2D variational theorem with respect to interparticle interaction of arbitrary shape was considered in the paper \cite{cherny01a}. Here we suggest the full form of the theorem in the low dimensions, which includes the variations of both one-particle dispersion and interaction. The theorem involves the Fourier transform of the scattering part of two-body wave function, which does not tend to zero at large distances in 1D nd 2D. Below we show how to overcome this difficulty with the help of generalized functions.

The standard approach to universal systems is to use the pseudopotential, proportional to the $\delta$-function. This scheme is quite convenient technically but leads to divergencies. Special methods are needed to get rid of the divergencies \cite{Combescot09,Valiente11,Valiente12a}. In the previous papers \cite{cherny00,cherny01a,cherny04}, we developed the approach, which is free from divergencies and applicable to short-range interactions of arbitrary shape.  Conceptually, the method of this paper is quite similar: we apply the variational theorem for an arbitrary potential and generalize this way the Tan adiabatic sweep theorem for one \cite{Barth11} and two \cite{werner12} dimensions, see Eq.~(\ref{Tanrelgen}). Equation (\ref{varchSh}) for the pair distribution function generalizes our previous results \cite{cherny00} to one and two dimensions.

The paper is organize as follows. In the next section, the variational theorem in arbitrary dimension is proved. In Sec.~\ref{GenLS}, the generalized Lippmann-Schwinger equation in 2D is written down explicitly. In order to verify the variational theorem with a specific example, we consider an exactly solvable case of the circle $\delta$-potential well. We apply the variational theorem to find the long-range momentum asymptotics of the average occupation numbers and short-range behaviour of the pair distribution function in Sec.~\ref{sec:Tanrel}. In Conclusion, the main results are discussed.

\section{The variational theorem for the scattering length in arbitrary dimension}
\label{sec:scat_leng}

Let us consider two particles of equal masses, interacting with the potential ${V}(r)$. It is assumed to be of the short-range type, that is, it decreases at infinity as $1/r^{\alpha}$ with the exponent $\alpha>3$ in 1D and 3D and $\alpha>2$ in 2D or faster. The dispersion relation is assumed to be of general form $T(p)$, which means that the operator of kinetic energy in the momentum representation is equal to $T(p)$. It is supposed to be a function of the absolute value of $\bm{p}$ and quite close to the usual free-particle dispersion at small momenta:
\begin{align}\label{Tprest}
T(p)-\frac{p^2}{2m}=O(p^4).
\end{align}
Here the symbol $O(z)$ denotes the terms of order $z$ or smaller. We also impose the condition $T(p)-\frac{p^2}{2m}=O(1/p^2)$ for large momenta.

The Hamiltonian of two particles takes the form $\hat{H}=T(p_{1})+T(p_{2})+{V}(|\bm{r}_{1}-\bm{r}_{2}|)$. As usual, it is convenient to introduce two pairs of the canonically conjugate variables: the relative momentum $\bm{p} =(\bm{p}_{1} -\bm{p}_{2})/2$ and position $\bm{r} =\bm{r}_{1} -\bm{r}_{2}$, and the total momentum $\bm{Q}=\bm{p}_{1}+\bm{p}_{2}$ and the center-of-mass position $\bm{R}=(\bm{r}_{1}+\bm{r}_{2})/2$. Then the kinetic energy of the two particles reads $T\left(\frac{\bm{Q}}{2} +\bm{p}\right) +T\left(\frac{\bm{Q}}{2}-\bm{p}\right)$. Although the center-of-mass motion no longer separates in general, the total momentum $\bm{Q} =\bm{p}_{1} +\bm{p}_{2}$ is still conserved, and thus it is a good quantum number. This allows us to look for  eigenfunctions in the form $\varPhi(\bm{r}_{1},\bm{r}_{2})=\exp(i\bm{Q}\cdot\bm{R}/\hbar)\varphi_{\bm{Q}}(\bm{r})$ with $\varphi_{\bm{Q}}$ being the wave function of the relative motion. In general, $\varphi_{\bm{Q}}(\bm{r})$ depends on the total momentum, while for the usual dispersion $p^2/(2m)$ it is independent of $\bm{Q}$.

We consider the two-particle scattering problem with zero total energy and momentum. Then kinetic energy is equal to $2T(p)$, and the two-body Schr\"odinger equation takes the form
\begin{equation}
2T\left(-i\hbar\nabla\right)\varphi(r)+{V}(r)\varphi(r)=0.
\label{twobody}
\end{equation}
For a short-range potential and free-particle dispersion obeying the restriction (\ref{Tprest}), the radially symmetric solution satisfies the boundary conditions for $r\to \infty$
\begin{equation}
\varphi(r)\simeq \begin{cases}
1-a/r, &D=3,\\
1-\ln r/\ln a, &D=2, \\
1-r/a, &D=1.
\end{cases}
\label{scatasymp}
\end{equation}
The parameter $a$ is called the scattering length. It can take arbitrary real value in 1D and 3D, while it is always positive in 2D \cite{cherny01a}. In two dimensions, we follow the definition given in Refs.~\cite{Verhaar84,cherny01a,cherny04,werner12}. At zero radius, the boundary condition is $|\varphi(r=0)|<\infty$ in 2D and 3D, while in 1D one can put \cite{jeszenszki18} $\left.\frac{\partial\varphi(r)}{\partial r}\right|_{r=0}=0$ when the potential is not very singular at $r=0$: $|V(r)|< C/r^s$ with  $s<1$.

It is convenient to separate the scattering part $\psi(r)$ of the wave function defined as
\begin{align}
\varphi(r)=1+\psi(r).
\label{psidef}
\end{align}
Multiplying Eq.~(\ref{twobody}) by $\varphi(r)$, using the identity $T\left(-i\hbar\nabla\right) 1=0$, and integrating over the whole real space yield
\begin{equation}
{U}^{}(0)
=\int{d}^Dr\left[\psi(r)2T\left(-i\hbar\nabla\right)\psi(r)
                            + {V}(r)[\varphi(r)]^2\right],
\label{33b}
\end{equation}
where $U(0)$ is the zero component of the scattering amplitude
\begin{align}
U^{}(p)=\int{d}^Dr\,{V}(r)\varphi(r)e^{-i\bm{p}\cdot\bm{r}}.\label{scatamp}
\end{align}

Equation (\ref{33b}) tells us that the zero-momentum scattering amplitude is a functional of the wave function and its stationary solution obeying the condition $\delta U(0)=0$ satisfies the Schr\"odinger equation (\ref{twobody}). It follows that when we vary $U(0)$ with respect to a parameter of the Hamiltonian at the stationary solution, we should not take into account a variation of the wave function. Then we arrive at the variational theorem \cite{cherny00} in arbitrary dimensions
\begin{align}
\delta{U}^{}(0)=\int{d}^D r
\Bigl[\psi(r)\delta T\left(-i\hbar\nabla\right)
\psi(r)
+\delta{V}(r)\varphi^{2}(r) \Bigr].\label{varth1}
\end{align}

It is convenient to write the first term in the Fourier representation
\begin{align}
\delta{U}^{}(0)=\int\frac{{d}^D p}{(2\pi)^D}\,
2\delta T({p})\psi^{2}(p)+\int{d}^D r\,\delta{V}(r)\varphi^{2}(r),
\label{varth}
\end{align}
where the Fourier transform of the scattering part of the wave function is given by
\begin{align}
\psi_{p}=-\frac{U(p)}{2T({p})}
\label{SchFour}
\end{align}
Equation (\ref{SchFour}) is nothing else but the Schr\"odinger equation (\ref{twobody}) in the Fourier representation.

Some caveats should be given here. In one and two dimensions, $\psi(r)$ does not tend to zero when $r\to\infty$ in accordance with Eq.~(\ref{scatasymp}). This corresponds to divergence of the Fourier transform of Eq.~(\ref{SchFour}) at zero momentum when $U(0)\not=0$. For this reason, its Fourier transform should be given in terms of generalized functions, and the appropriate regularization should be done \cite{Gelfand:book64}. The singular part of $\psi_{p}$ is proportional to $1/p^2$, and the Fourier transform of $1/p^2$
can be understood as the solution of equation $\nabla^2 f(r)=-\delta(\bm{r})$ with $\delta(\bm{r})$ being the Dirac $\delta$ function in $D$ dimension. Then the function $f(r)$ is a sum of a particular solution and arbitrary solutions of the homogeneous equation $\nabla^2 f(r)=0$, and we get an indeterminacy due to the regularization. However, the indeterminacy can be eliminated by
the boundary condition (\ref{scatasymp}). We obtain $f(r)=-r/2$, $f(r)=-\ln r/(2\pi)$, and $f(r)=1/(4\pi r)$ in one, two, and three dimensions, respectively. This reasoning is analogous to that of used for constructing Green functions in the Lippmann-Schwinger equation.

It follows from the above considerations that the main singularity $-U(0)m/(\hbar^2 p^2)$ at $p=0$ in the r.h.s. of Eq.~(\ref{SchFour}) determines the main asymptotics of $\psi(r)$ for $r\to\infty$. Using the definition of the scattering length (\ref{scatasymp}), we arrive at the relation between the scattering length and the scattering amplitude at zero momentum:
\begin{align}
U^{}(0)=&\frac{\hbar^{2}}{m}\times\begin{cases}
4\pi a, &D=3,\\
-2\pi/\ln a,&D=2,\\
-2/a,&D=1.
\end{cases}
\label{adef}
\end{align}

The variation of the scattering amplitude with respect to the interaction potential in two and three dimensions was considered in Refs.~\cite{cherny01a} and \cite{Popov79}, respectively. The variation with the both one-particle dispersion and interaction potential in three dimension was found in Ref.~\cite{cherny00}.

Finally, the variational theorem (\ref{varth}) in arbitrary dimension reads
\begin{align}
  \frac{\delta U(0)}{\delta T({p})} =&\frac{1}{(2\pi)^D}\frac{U^{2}(p)}{2T^{2}({p})}
  =\frac{m^2U^{2}(p)}{2^{D-1}\pi^D\hbar^4p^4}, \label{varTp}\\
  \frac{\delta U(0)}{\delta V(r)} =&\varphi^{2}(r).\label{varV}
\end{align}
The second equality in Eq.~(\ref{varTp}) assumes that the variation is taken in the vicinity of the usual free-particle dispersion.

The short-range potential $V(r)$ is supposed to be localized within its characteristic radius $r_0$. Then the asymptotics (\ref{scatasymp}) is realized for $r\gg r_0$, and, besides,   $U(p)\simeq U(0)$ when $p\ll 1/r_0$.

\section{The generalized  Lippmann-Schwinger equation for arbitrary one-particle dispersion in two dimensions}
\label{GenLS}

The considerations of Sec.~\ref{sec:scat_leng} allow us to correctly define the corresponding form of the Lippmann-Schwinger equation for arbitrary one-particle dispersion. Equation (\ref{SchFour}) implies that $\psi(r)$ is the convolution of $V(r)\varphi(r)$ and the Green function, which is the Fourier transform of $1/T(p)$ with arbitrary $T(p)$. In this section, we consider as an example the generalized Lippmann-Schwinger equation in two dimensions and its exact analytical solution for a specific interaction potential.

Separating the singular part $2m/(\hbar^2p^2)$ in $\psi_{p}$ (\ref{SchFour}), taking the Fourier transformation, and integrating over the polar angle, we obtain with the help of Eqs.~(\ref{psidef}) and (\ref{scatamp})
\begin{align}
\varphi(r)=&1+\frac{m}{\hbar^2}\int_{0}^{\infty} dy\,y V(y)\varphi(y)\Big[\theta(r-y)\ln r\nonumber\\
&+\theta(y-r)\ln y+I(r,y)\Big],\label{LS2D}\\
I(r,y)=&\frac{\hbar^2}{2m}\int_{0}^{\infty} dp\,p\left(\frac{2m}{\hbar^2p^2}-\frac{1}{T(p)}\right)J_{0}(pr)J_{0}(py),\label{Iry}
\end{align}
where $\theta(z)$ and $J_{0}(z)$ are the Heaviside step function and Bessel function of zero order, respectively. Due to the condition (\ref{Tprest}),
the integrand in Eq.~(\ref{Iry}) is regular at zero momentum.

The generalized Lippmann-Schwinger equation (\ref{LS2D}) admits an analytical solution for the interaction potential
\begin{align}\label{Vdelt}
V(r)=\frac{\hbar^2}{mr_0\varkappa}\delta(r-r_0).
\end{align}
Here the dimensionless control parameter $\varkappa$ determines the strength of interaction; it can be positive or negative. The presence of the $\delta$ function enables us to get rid of the integral, and the integral equation (\ref{LS2D}) becomes algebraic. Substituting Eq.~(\ref{Vdelt}) into the generalized Lippmann-Schwinger equation, we obtain the solution for $\varphi(r_0)$
\begin{align}\label{phir0}
\varphi(r_0)=\frac{\varkappa}{\varkappa-\ln r_0- I(r_0,r_0)}.
\end{align}
Once $\varphi(r_0)$ is known, we find the scattering amplitude for arbitrary one-particle dispersion with the definition (\ref{scatamp})
\begin{align}
U(p)=&U(0)J_{0}(pr_0), \label{Uparb}\\
U(0)=&\frac{2\pi\hbar^2}{m\varkappa}\varphi(r_0). \label{Uparb0}
\end{align}

Now the scattering amplitude of the standard Schr\"odinger equation (\ref{twobody}) is obtained from Eqs.~(\ref{phir0}), (\ref{Uparb}), and (\ref{Uparb0}) in the particular case $T(p)=\hbar^2p^2/(2m)$, which corresponds to $I(r_0,r_0)=0$. With the help of Eq.~(\ref{adef}), we get the scattering length
\begin{align}\label{a2D}
a=r_0e^{-\varkappa}.
\end{align}

It is not difficult to find explicitly the variation $\delta U(0)$ with respect to $\delta T(p)$ and directly check the variation theorem (\ref{varTp}) in the particular case of interaction (\ref{Vdelt}), see Appendix \ref{Tpcheck}.

\begin{figure}[!tbp]
\begin{center}
\includegraphics[width=\columnwidth,clip=true]{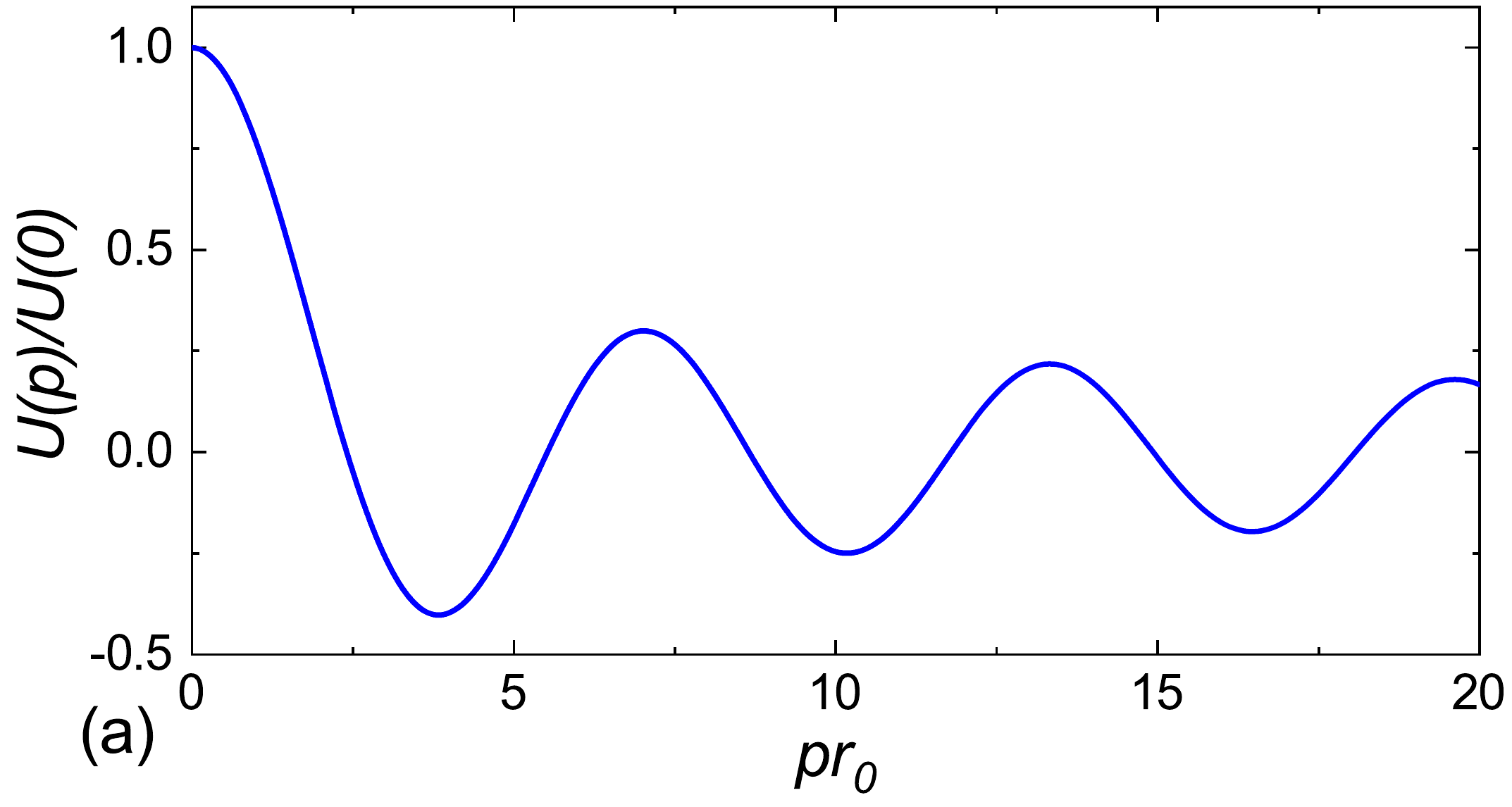}\\
\includegraphics[width=\columnwidth,clip=true]{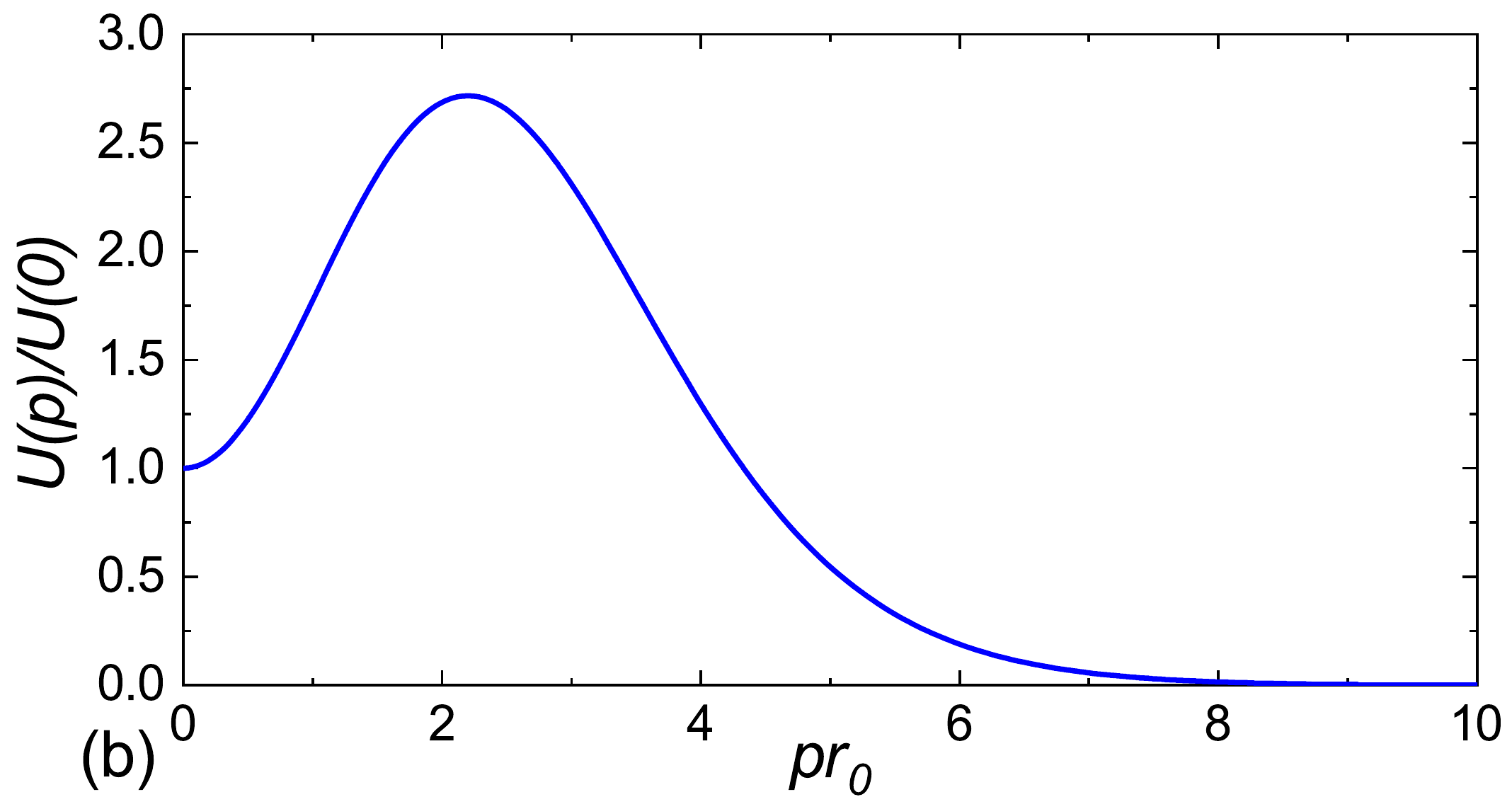}\\
\end{center}
\caption{\label{fig:Up} The scattering amplitude (\ref{scatamp}) (in units of $U(0)={-2\pi\hbar^2}/{m\ln a}$) as a function of the wavenumber (in units of the inverse potential radius $1/r_0$) for (a) the circle $\delta$-potential well (\ref{Vdelt}) and (b) the Gaussian potential (\ref{Gpot}) with the control parameter $\alpha=-9$. The approximation $U(p)\simeq U(0)$ is satisfied only when $p\ll 1/r_0$, and the scattering amplitude $U(p)$ always falls off to zero at large wave vectors $p\gg 1/r_0$.
}
\end{figure}

The behaviour of the scattering amplitude for the potential (\ref{Vdelt}) is shown in Fig.~\ref{fig:Up}a. The ratio $U(p)/U(0)$ is the Bessel function of zero order in accordance with Eq.~(\ref{Uparb}). We also calculate numerically the scattering amplitude (\ref{scatamp}) for the Gaussian potential
\begin{align}\label{Gpot}
V(r)=\frac{\hbar^2\alpha}{m r_0^2}\exp\left(-\frac{r^2}{r_0^2}\right),
\end{align}
see the details in Ref.~\cite{jeszenszki18}. Here $\alpha$ is the dimensionless control parameter. We choose $\alpha=-9$, for which $a/r_0=0.3289\ldots$. The results are represented in Fig.~\ref{fig:Up}b. The approximation $U(p)\simeq U(0)$ is satisfied only when $p r_0\ll 1$, and at large wave vectors $p\gg1/r_{0}$, $U(p)$ tends to zero, which is a typical behaviour for short-range potentials.

\section{The Tan adiabatic sweep theorem and the pair distribution function in one and two dimensions}
\label{sec:Tanrel}

The Tan adiabatic sweep theorem \cite{Tan2008b,Tan2008a} relates the derivative of a thermodynamic potential of a universal system with respect to the scattering length to the asymptotic behaviour of the average occupation numbers for large momenta. In our previous paper \cite{cherny21}, a simple derivation of the Tan theorem is given for three-dimensional universal systems. Here we consider the ground state energy of a homogeneous system of interacting spinless bosons in one and two dimensions.

In general, the ground state energy per $D$-volume can be written down through the average occupation numbers $n_{p}$ and pair distribution function $g(r)$
\begin{align}
\varepsilon =\frac{E}{L^D}=\frac{1}{(2\pi)^D}\int{d}^Dp\, T(p)n_{p} +\frac{n^2}{2}\int{d}^Dr\,g(r)V(r)
\label{GSE}
\end{align}
with $L$ and $n=N/L^D$ being the system size and density of particles, respectively. We emphasize that the physical dimensions of the energy per $D$-volume are different in various dimensions; the same relates to the density of particles.

In general, both $n_{p}$ and $g(r)$ depend on the one-particle dispersion and interaction. However, in agreement with the Hellmann-Feynman theorem, one can take into consideration only the explicit dependence on the parameters while varying the energy (see the detailed discussion in Ref.~\cite{cherny21}):
\begin{align}
\frac{\delta \varepsilon}{\delta T(p)}= &\, \frac{n_{p}}{(2\pi)^D}, \label{np}\\
\frac{\delta \varepsilon}{\delta V(r)}= &\, \frac{n^2}{2}g(r).\label{gr}
\end{align}

\subsection{The long-range asymptotics of the occupation numbers}
\label{npas}

We find the momentum distribution at high momenta. The scattering amplitude $U(p)$ is supposed to be found from the ordinary Schr\"odinger equation (\ref{twobody}) with the usual dispersion relation.

Technically, it is convenient to hold the relation (\ref{adef}) between $U(0)$ and $a$ while varying the dispersion relation. Then we vary the one-particle dispersion $T(p)$ near $T(p)=\hbar^2p^2/(2m)$ but keep the mass constant. For universal systems, the variation of the ground-state energy depends on $\delta T(p)$ only through the variation of the scattering length: $\frac{\delta \varepsilon}{\delta T(p)}=\frac{\partial \varepsilon}{\partial a}\frac{\delta a}{\delta T(p)}$. Equating this expression and Eq.~(\ref{np}), and using Eq.~(\ref{adef}) and the variation theorem (\ref{varTp}), we arrive at the relations
\begin{align}\label{Tanrelgen}
p^4 n_{p}=\frac{4 m}{\hbar^2}\frac{\partial\varepsilon}{\partial a}\frac{U^{2}(p)}{U^{2}(0)}\times\begin{cases}
\pi a, &D=2,\\
1,&D=1.
\end{cases}
\end{align}
where $U(p)$ is the scattering amplitude (\ref{scatamp}) for the standard Sch\"odinger equation (\ref{twobody}) with the usual dispersion $T(-i\hbar\nabla) =-\hbar^2\nabla^2/(2m)$.

Thus we obtain the generalization of Tan's adiabatic sweep theorem for arbitrary short-range potentials in 1D and 2D. Equation (\ref{Tanrelgen}) is valid for arbitrary $p\gg 1/\xi$, where $\xi$ is a parameter with the dimension of length, which determines the characteristic scale of the many-body effects. Within the mean-field approximation, $\xi=\hbar/\sqrt{\mu m}$ is the healing length with $\mu$ being the chemical potential, and the approximation is applicable when $\xi\gg n^{1/D}$. In the dilute 2D Bose gas, when the gas parameter $na^2$ is small, we have \cite{schick71,cherny01a,mora09} $\xi\simeq\frac{1}{2}\sqrt{\frac{\ln na^2}{-\pi n}}$. In one dimension, the healing length is given by \cite{lieb63:1} $\xi\simeq\sqrt{\frac{-a}{2n}}$, and the mean-field approach works when the density is sufficiently big: $\sqrt{-na} \gg 1$. For large $na^2$ in the 2D case, the ground state becomes unstable against cluster formation and further crystallization \cite{Xing90,Pilati05}. For small $-na$ in 1D case, the parameter $\xi$ is of order of mean distance between particles: $\xi\sim 1/n$. Within the range $1/\xi \ll p\ll 1/r_0$, the scattering amplitude $U(p)$ is almost constant: $U(p)\simeq U(0)$, see the discussion in Sec.~\ref{sec:scat_leng} above. Then we arrive at Tan's theorem \cite{Tan2008b}
\begin{align}\label{Tanrel}
p^4 n_{p}\simeq{\cal C}=\frac{4 m}{\hbar^2}\frac{\partial\varepsilon}{\partial a}\times\begin{cases}
\pi a, &D=2,\\
1,&D=1,
\end{cases}
\end{align}
where ${\cal C}$ is the parameter introduced by Tan \cite{Tan2008a} and called Tan's contact. Equation (\ref{Tanrel}) can formally be derived within the pseudopotential approach, where the relation $U(p)= U(0)$ is exact. For a ``usual" potential of arbitrary shape, the limit $p^4 n_{p}$ is equal to zero when $p\to\infty$, because $U(p)$ tends to zero at large momenta, see Fig.~\ref{fig:Up}.

\subsection{The short-range spatial correlations}
\label{grsmallr}

The short-range spatial correlations for a universal system are obtained in the same manner \cite{cherny00,cherny01,cherny21}. We find from Eq.~(\ref{gr}) that $\frac{\delta \varepsilon}{\delta V(r)} =\frac{n^2}{2}g(r)=\frac{\partial \varepsilon }{\partial a}\frac{\delta a}{\delta V(r)}$. Using the relation (\ref{adef}) and variational theorem (\ref{varV}) finally yields
\begin{align}
g(r)=\frac{m}{\hbar^2n^2}\frac{\partial \varepsilon }{\partial a}\varphi^2(r)\times\begin{cases}
a(\ln a)^2/\pi , &D=2,\\
a^2,&D=1.
\end{cases} \label{varchSh}
\end{align}
As one can see, at short distances the pair distribution function is proportional \cite{cherny00,cherny01a} to the squared wave function of the standard  Schrodinger equation (\ref{twobody}). We note that in two dimension, the renormalized  wave function $-\varphi(r)\ln a$ obeys to the boundary condition $-\varphi(r)\ln a\simeq \ln(r/a)$ in accordance with Eq.~(\ref{scatasymp}). The prefactor can be written in terms of Tan's contact (\ref{Tanrel})
\begin{align}
g(r)=\frac{{\cal C}}{4n^2}\varphi^2(r)\times\begin{cases}
(\ln a)^2/\pi^2 , &D=2,\\
a^2,&D=1.
\end{cases} \label{varVrTan}
\end{align}

Equations (\ref{varchSh}) and (\ref{varVrTan}) are valid at short distances $r\ll \xi$. It follows from the asymptotics (\ref{scatasymp}) that in two dimensions, $\ln^2a\,\varphi^2(r)$ is approximately equal to $\ln^2(r/a)$ for $r_0\ll r\ll \xi$. In one dimension, $a^2\varphi^2(r)\simeq(a-r)^2$ under the same conditions. Then the relation (\ref{varVrTan}) leads to
\begin{align}
g(r)=\frac{{\cal C}}{4n^2}\times\begin{cases}
\ln^2(r/a)/\pi^2 , &D=2,\\
(a-r)^2,&D=1.
\end{cases}
 \label{grps}
\end{align}
within the range $r_0 \ll r\ll \xi$. These  results are in consistency with Refs.~\cite{werner12,Barth11} obtained in the framework of the pseudopotential approach.

Note that Eqs.~(\ref{Tanrelgen}) and (\ref{varchSh}) are also valid in the canonical and grand canonical ensembles if the derivative of the energy is replaced by the derivative of the corresponding thermodynamic potential, see the discussion in Ref.~\cite{cherny21}.

\section{Conclusion}
\label{sec:concl}

We generalize the full form of the variational theorem for the scattering length \cite{cherny00} to arbitrary dimension, where the variation is taken with respect to one-particle dispersion (\ref{varTp}) and interaction potential (\ref{varV}), and verify it in the particular case of the circle $\delta$-potential well (\ref{Vdelt}), see Appendix \ref{Tpcheck}.

With the help of the variational theorem, the Tan adiabatic sweep theorem is extended to short-range potentials of arbitrary shape in 1D and 2D for a universal system of spinless bosons, see Eq.~(\ref{Tanrelgen}). This equation gives us the long-range asymptotics of the mean occupation numbers, which contains the additional factor proportional to the squared scattering amplitude (\ref{scatamp}) for the standard Schr\"odinger equation (\ref{twobody}) with the dispersion $T(-i\hbar\nabla) =-\hbar^2\nabla^2/(2m)$. Equation (\ref{Tanrelgen}) is applicable for wave vectors more than $1/\xi$, where $\xi$ is the characteristic scale of the many-body effects (see the discussion in Sec.~\ref{npas}). By contrast, the Tan relation (\ref{Tanrel}) is valid for $1/\xi \ll p \ll 1/r_{0}$, where $r_{0}$ is the radius of the interaction potential.  For the non-polarized two-component fermions, the analogous equations can also be written down with the same method.

The short-range behaviour of the pair distribution function (\ref{varchSh}) is obtained  by means of the variation with respect to interaction potential (\ref{varV}). The pair distribution function is known at all distances $r\ll\xi$, while the pseudopotential approach leads to Eq.~(\ref{grps}), which described $g(r)$  beyond the radius of interaction potential $r_0\ll r\ll\xi$.

As a prospect, one can calculate the mean kinetic and interaction energies for a homogeneous universal many-body systems in the low dimensions by analogy with the paper \cite{cherny21}.

\section{Acknowledgement}

The author acknowledges support from the JINR--IFIN-HH projects.

\appendix

\section{The variational theorem for the circle delta-potential well}
\label{Tpcheck}

In this appendix, we prove straightforwardly the variational theorem (\ref{varTp}) in two dimensions for the potential (\ref{Vdelt}). As in Sec.~\ref{npas}, it is technically convenient to keep the mass constant to hold the relation (\ref{adef}) between $U(0)$ and $a$.

We obtain from Eqs.~(\ref{phir0}) and (\ref{Uparb0})
\begin{align}\label{delU0delTp}
  \frac{\delta U(0)}{\delta T(p)}=\frac{2\pi\hbar^2}{m\varkappa}\frac{\delta \varphi(r_0)}{\delta T(p)}
  =\frac{2\pi\hbar^2}{m\varkappa^2} \varphi^2(r_0)\frac{\delta I(r_0,r_0)}{\delta T(p)}.
\end{align}
It follows from the definition (\ref{Iry}) that
\begin{align}\label{delIdelTp}
\frac{\delta I(r_0,r_0)}{\delta T(p)}= \frac{\hbar^2}{2m}\frac{J_{0}^2(pr_0)}{2\pi T^2(p)},
\end{align}
because the volume element in two dimensions is given by $d^2p=2\pi p\, dp$. After substituting Eq.~(\ref{delIdelTp}) into Eq.~(\ref{delU0delTp}), we are left with
\begin{align}\label{delU0delTp1}
 \frac{\delta U(0)}{\delta T(p)}= \frac{\hbar^4}{m^2\varkappa^2}\frac{\varphi^2(r_0)J_{0}^2(pr_0)}{2 T^2(p)},
\end{align}
which coincides with the relation (\ref{varTp}) for $D=2$ by virtue of Eqs.~(\ref{Uparb}) and (\ref{Uparb0}).

\bibliography{tanvar}

\end{document}